\documentclass[%
 reprint,
superscriptaddress,
 amsmath,amssymb,
 aps,prl,showpacs
]{revtex4-1}

\usepackage{graphicx}
\usepackage{dcolumn}
\usepackage{bm}
\usepackage{natbib} 

\begin{document}

\preprint{APS/123-QED}

\title{Signatures of pair-density wave order in phase-sensitive measurements of La$_{2-x}$Ba$_x$CuO$_4$-Nb Josephson junctions and SQUIDs}

\author{D.R. Hamilton}
\affiliation{%
 Department of Physics, University of Illinois at Urbana-Champaign, Urbana, Illinois 61801, USA
}%
\author{G.D. Gu}
\affiliation{%
 Brookhaven National Laboratory, Upton, New York 11973-5000, USA
}%
\author{E. Fradkin}
\affiliation{%
 Department of Physics, University of Illinois at Urbana-Champaign, Urbana, Illinois 61801, USA
}%
\author{D.J. Van Harlingen}%
\affiliation{%
 Department of Physics, University of Illinois at Urbana-Champaign, Urbana, Illinois 61801, USA
}%

\date{\today}

\begin{abstract}
The interplay of charge order, spin order, and superconductivity in La$_{2-x}$Ba$_x$CuO$_4$ creates a complex physical system that hosts several interesting phases, such as two-dimensional superconductivity within the CuO$_2$ planes and the ordered pair-density wave state in which charge ordering is intertwined with superconductivity. Using Josephson interferometry techniques, we measure the current-phase relation of junctions and SQUIDs incorporating this material and observe a significant sin($2\phi$)-component indicative of closely-spaced alternations of the sign of the Josephson coupling predicted by the pair-density wave model. We find that the ratio of the sin(2$\phi$)-component to the conventional sin($\phi$)-component to be largest near x=1/8 doping, where the pair-density wave state is believed to be the strongest, and that it increases with increasing temperature as the Josephson coupling in the junction weakens.
\end{abstract}

\maketitle

\begin{figure}
    \includegraphics[width=\columnwidth]{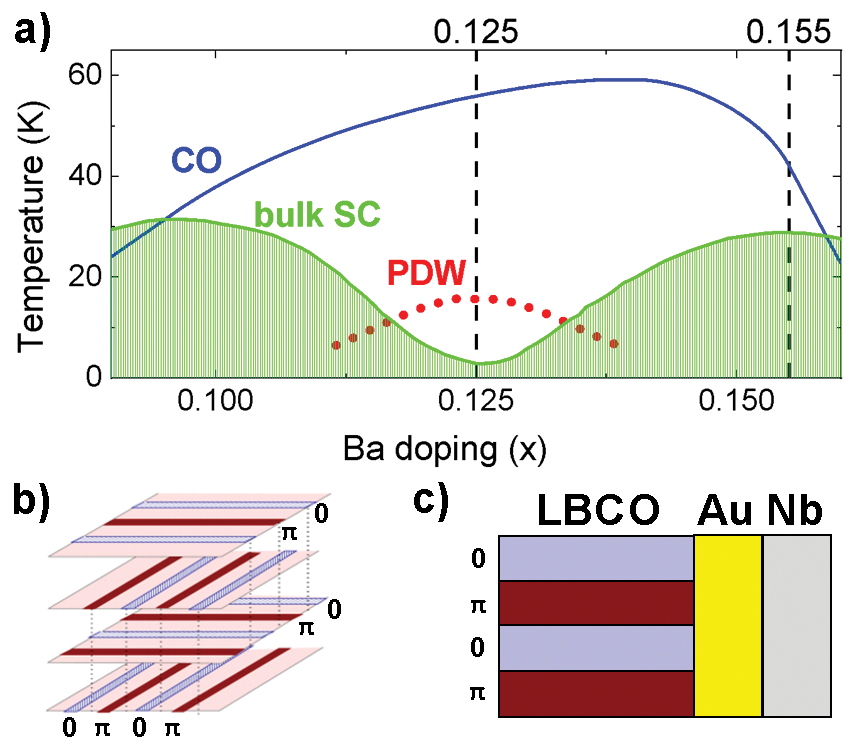}
    \caption{(a) Phase diagram of La$_{2-x}$Ba$_x$CuO$_4$, reproduced from \cite{hucker_stripe_2011}. Dotted line curve indicates estimated onset of pair-density wave state. T$_c$ is suppressed at $x=1/8$ doping, which coincides with charge and spin ordering. Dashed verical lines at $x=1/8$ and $x=0.155$ indicate doping of crystals used in our experiment. (b) In the PDW model, adjacent charge stripes experience a phase shift of $\pi$ in the superconducting order, and adjacent planes are rotated by $90^{\circ}$. Reproduced from \cite{berg_dynamical_2007}. (c) Schematic of a Josephson junction formed on an LBCO superconductor, formed over several stripes with alternating $0$ and $\pi$ coupling.}  
    \label{background}
    \end{figure}

La$_{2-x}$Ba$_x$CuO$_4$ (LBCO), the first discovered high-temperature superconductor \cite{bednorz_possible_1986}, exhibits unusual properties near $x=1/8$ doping, including a dramatic suppression in T$_c$ \cite{moodenbaugh_superconducting_1988},  charge and spin stripe ordering \cite{fujita_stripe_2004,hucker_stripe_2011},  and a frustration of Josephson coupling between CuO$_2$ planes leading to 2D superconductivity \cite{li_two-dimensional_2007}. Much of the anomalous behavior of La$_{2-x}$Ba$_x$CuO$_4$ can be explained by the existence of a pair-density wave (PDW) state, in which the sign of the superconducting order parameter varies periodically with position \cite{berg_theory_2009}. 
Experiments using nonlinear terahertz optics and STM appear to show some evidence of this state in LBCO and Bi$_2$Sr$_2$CaCu$_2$O$_8$, respectively \cite{rajasekaran_probing_2018,hamidian_detection_2016,edkins_magnetic-field_2018}. In the PDW model, the superconducting order on adjacent charge stripes is phase-shifted by $\pi$, and the charge density wave order rotates by $90^{\circ}$ between adjacent CuO$_2$ planes, leading to a cancellation of the effective Josephson coupling between each CuO$_2$ plane and its three nearest neighbors (see Fig. \ref{background}b) \cite{berg_dynamical_2007}.  Thermal melting of such a state is expected to yield multiple interesting phases, most notably a charge-$4e$ superconducting condensate \cite{berg_charge-4e_2009}.  The presence of such a state in LBCO would manifest experimentally as a flux periodicity of $\Phi_0/2 = hc/4e$ in a SQUID having junctions between a conventional charge-$2e$ superconductor and LBCO \cite{berg_charge-4e_2009}. More specifically, we expect that the rapid spatially-modulated sign changes in the Josephson coupling that arise from a pair-density wave state suppress the first-order Josephson coupling and manifest itself as a significant sin(2$\phi$) harmonic in the the current-phase relation of a junction containing LBCO (see Fig. \ref{background}c).  This phenomena has been predicted and observed in other junctions with spatially alternating critical current density  \cite{buzdin_periodic_2003,moshe_shapiro_2007,stoutimore_second-harmonic_2018,schneider_half-h/2e_2004}. Additionally, we expect the fraction of Josephson current exhibiting a sin(2$\phi$) current-phase relation to increase with T as the interlayer Josephson coupling and conventional 3D superconductivity are suppressed within LBCO, giving way to an increasing proportion of spatially varying 2D superconductivity within the CuO$_2$ planes \cite{berg_dynamical_2007}.

    In this Letter, we present measurements of Josephson junctions and SQUIDs fabricated onto crystals of La$_{2-x}$Ba$_x$CuO$_4$ at a range of temperatures and dopings that show evidence for the onset of a PDW state. We observe a sin(2$\phi$) component of the Josephson current-phase relation of the devices near $x=1/8$ doping, corresponding to $\Phi_0/2
   $ magnetic flux periodicity. The proportion of sin(2$\phi$) to conventional sin($\phi$) increases with temperature as 3D superconductivity in the crystal is suppressed, reaching above 25\% in the highest temperatures at which we can measure the supercurrent. In devices measured far from $x=1/8$ doping, the Josephson current-phase relation shows only a very small sin(2$\phi$) component, less than 5\%. In these experiments, we also were able to measure the order parameter symmetry of the superconductivity in an LBCO crystal, and find evidence for $d-$wave pairing symmetry, as observed in all other cuprate superconductors. \cite{wollman_experimental_1993,tsuei_determination_1997}.
    
	Measurements were performed on devices fabricated onto La$_{2-x}$Ba$_x$CuO$_4$ crystals grown at the nominal dopings of $x=0.125$ at the minimum T$_c$, slightly-underdoped at $x=0.120$, and overdoped near the maximum T$_c$ at $x=0.155$. In this paper, we will primarily discuss what we observed at $x=0.125$ and $x=0.155$, as these dopings were measured more extensively. The LBCO crystals were grown using a floating-zone technique. Susceptibility measurements indicate T$_c$'s of 7K, 11K, and 30K for the $x=0.125$, $x=0.120$ and $x=0.155$ crystals, respectively.  
    
    Each crystal is oriented using a Laue x-ray camera and polished to create smooth facets on the a, b, and c faces. The locations of junctions on the a and b faces of the crystal are defined by masking the edges with a polymer resist, ion-milling to clean the LBCO surface, and depositing $50-100nm$ of Au by electron beam evaporation.  To improve contact between the Au barrier and the LBCO, the crystal is annealed at $400^{\circ}C$ in an O$_2$ atmosphere for $3-4$ hours. After annealing, the crystal is attached to a Si chip using a droplet of polyimide as an adhesive, with the crystal $c$-axis oriented perpendicular to the substrate. A second mask applied over the edge of the crystal at the location of the barriers defines the Nb superconducting electrodes, which are deposited by sputtering to form a smooth connection along the substrate and onto the edge of the crystal. Measurements were performed in both a $^4$He cryostat with a base temperature of $1K$, and in a single-shot $^3$He system with a base temperature of $310mK$. As typical for S-N-S junctions, the resulting Nb-Au-LBCO junctions exhibit an onset of supercurrents at temperatures well below the bulk crystal T$_c$.  As a result, the measurements we report were made at temperatures well below T$_c$ to obtain measurable critical currents ($>$ 1nA).

\begin{figure}
    \includegraphics[width=\columnwidth]{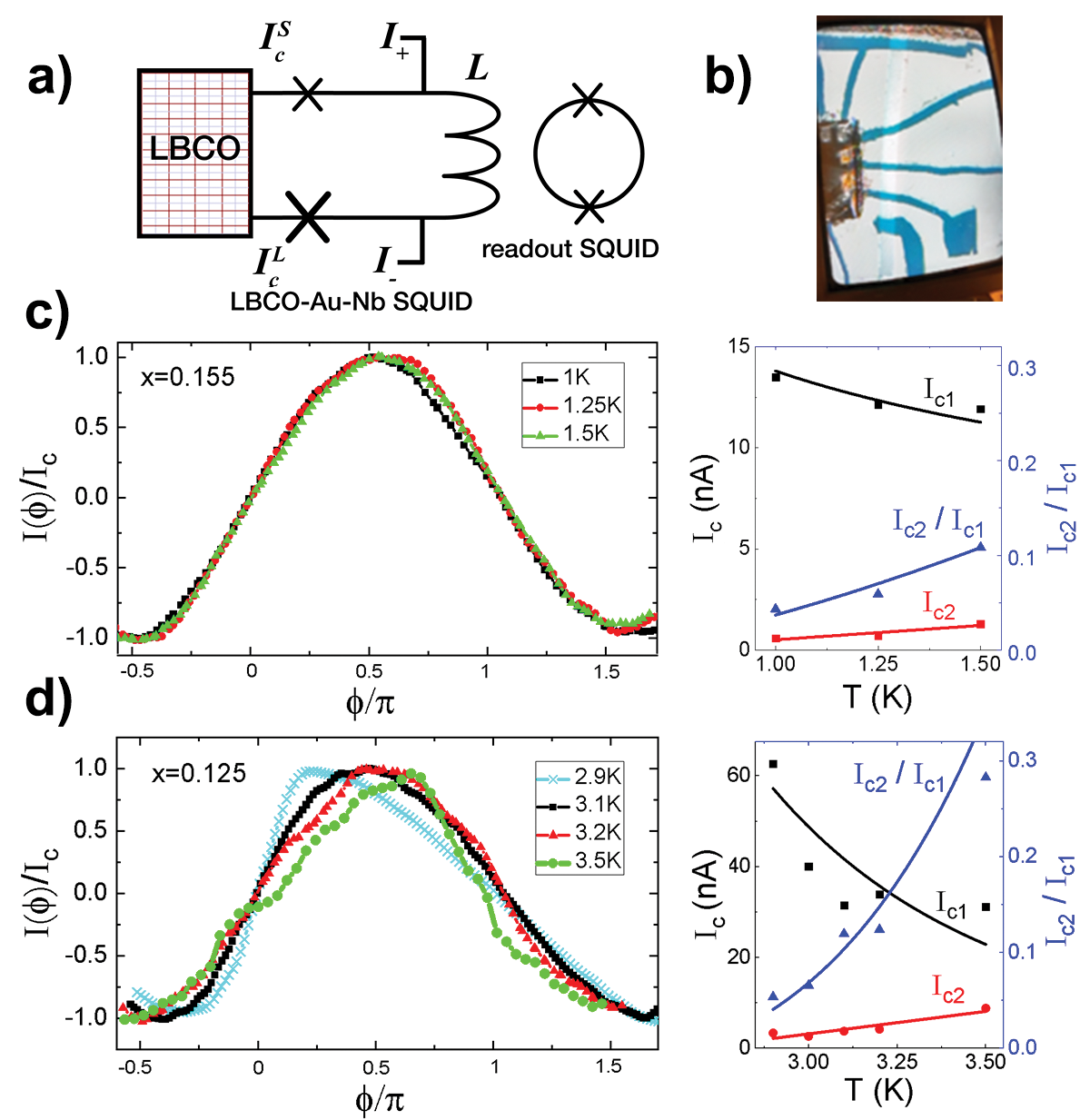}
    \caption{(a) Schematic of the current-phase relation (CPR) measurement on an La$_{2-x}$Ba$_x$CuO$_4$ crystal SQUID. (b) Optical microscope image of crystal device, containing an on-chip inductance for the CPR measurement. (c) Normalized CPR measurements vs T at $x=0.155$ doping (left), and Fourier analysis of the CPR vs T (right). Points in Fourier graph are measured amplitudes (or ratios), curves are fits to theoretical temperature dependence. d) CPR and Fourier analysis at $x=1/8$ doping. This sample sees more deviation from typical sin($\phi$) behavior, compared to the $x=0.155$ sample. The ratio of $2^{nd}/1^{st}$ harmonic is greatest in the $x=1/8$ sample over the range of measurable temperatures.}
    \label{directcpr}
    \end{figure}

	To make a direct measurement of the current-phase relation of a junction, the LBCO-Au-Nb Josephson junction of interest is fabricated in parallel with a superconducting inductor $L$ (see Fig. \ref{directcpr}a). An applied current $I$ divides between the junction path through the LBCO crystal and through the inductor to maintain equal phase differences.  In practice, since annealed Au is necessary for good electrical contact between Nb and LBCO, the junction path always contains two junctions in series.  However, by making the second contact much larger so that its critical current $I_c^L >> I_c^S$, the critical current of the smaller junction, we can ensure that the phase difference across the large junction is negligible.  We also keep the path length through the junction short to ensure that the the dominant phase difference is the phase drop $\phi$ across the small junction, allowing us to directly measure its current-phase relation $I_J(\phi)$. Under these conditions, the current divides so that $I=I_{J}(\phi)+(\Phi/2\pi L)\phi$, where $\Phi$ is the induced flux in the inductor.  Using a flux transformer to couple the inductor to a commercial SQUID, we can measure $\Phi$ in the loop as a function of bias current $I$, and obtain directly the current-phase relation $I_{J}(\phi)$\cite{english_observation_2016}.  For the full current-phase relation to be extracted with this method, we require $\phi_{J}$ to be single-valued.  Hence, the constraint $\beta_L=2\pi$$LI_c/\Phi_0<1$ must be satisfied in order for our specific sample geometry to yield a single-valued CPR. 

Normalized CPR curves measured for $x=0.155$ and $x=1/8$ samples over a range of temperatures are shown in Fig. \ref{directcpr}c and \ref{directcpr}d, respectively. As $T$ increases, the CPR of the $x=1/8$ sample becomes noticeably more forward skewed, and increasingly deviates from the typical sin($\phi$) CPR of a typical Josephson junction. In contrast, the CPR of the $x=0.155$ sample appears dominated by the conventional sin($\phi$) component of the current-phase relation over the temperatures measured, though a slight forward skewness is present.

Performing a Fourier analysis of the measured CPR curves, we can see in Fig. \ref{directcpr}c and Fig. \ref{directcpr}d that both samples see a decrease in $I_{c1}$, the sin($\phi$) component of the current phase relation, with temperature. These curves fit theoretical predictions for the temperature dependence of the critical current in diffusive S-N-S junctions \cite{likharev_superconducting_1979}. In contrast, $I_{c2}$, the sin(2$\phi$) component, increases roughly linearly with temperature in both samples, with the ratio of $I_{c2}$/$I_{c1}$ significantly larger in the $x=1/8$ sample compared to the $x=0.155$ sample.		
    
While susceptibility measurements show predictably that our $x=0.155$ crystals have a higher T$_c$ than the $x=1/8$ crystals, the critical currents of the junction on the $x=0.155$ crystal were considerably smaller, requiring measurement at lower temperatures to reach an acceptable signal-to-noise ratio. We attribute this difference to sample fabrication variations since we annealed the $x=1/8$ crystal for a longer time, causing the Au to diffuse a longer distance through the poorly conducting surface layers of the $x=1/8$ crystal compared to the $x=0.155$ crystal, thus, improving coupling between the LBCO crystal and Nb film. 
 
At $T=2.9K$ and below for the $x=1/8$ sample, we observe a slight negative skewness of the critical current.  At these temperatures, the critical current is large enough for the condition $\beta_L=2\pi$$LI_c/\Phi_0<1$ to no longer be satisfied, causing hysteretic switching to take place in the SQUID circuit. When this hysteresis is combined with noise-rounding, the measured CPR will appear negatively skewed, as described in \cite{english_observation_2016}. 
\begin{figure}
    \includegraphics[width=\columnwidth]{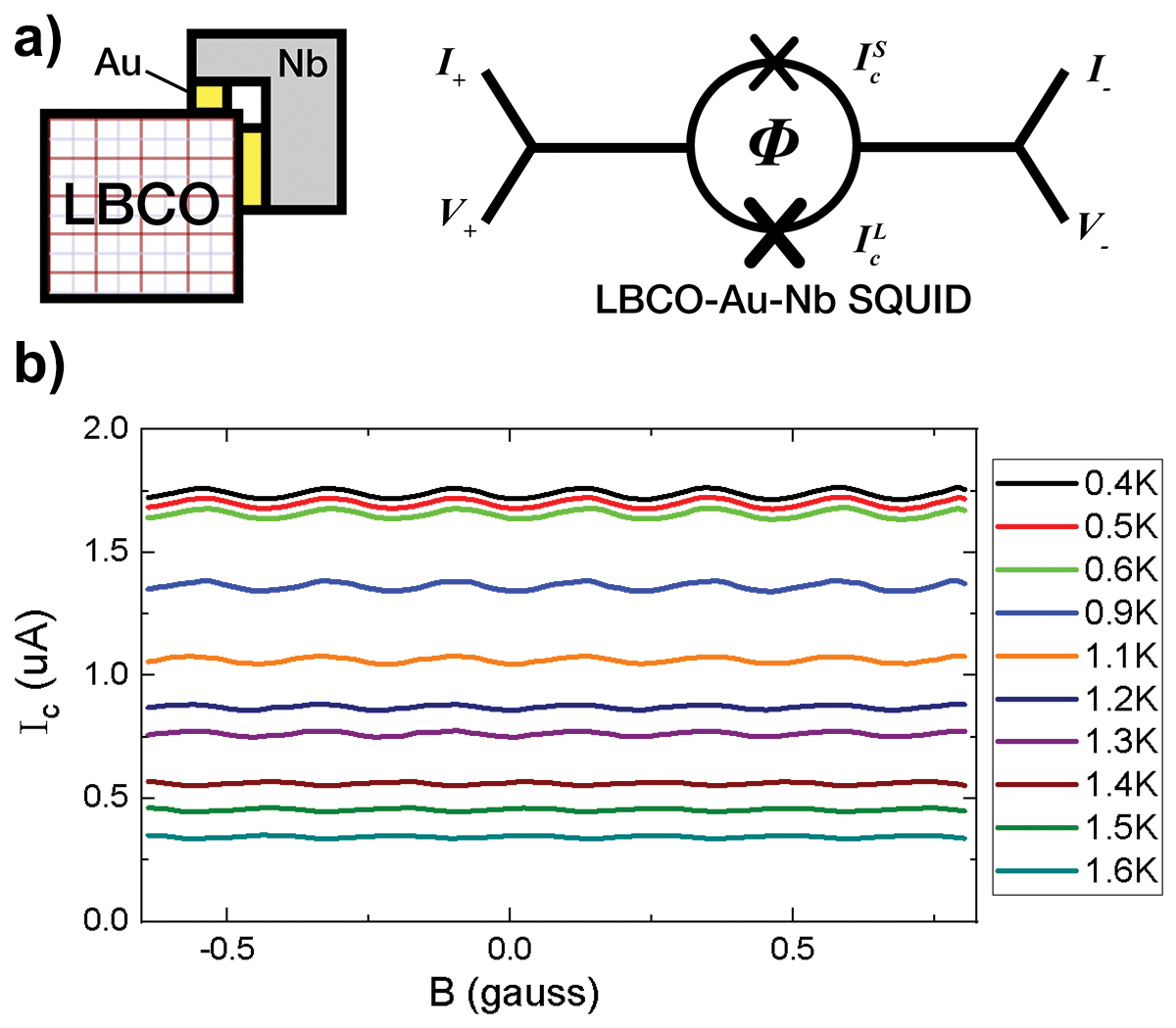}
    \caption{(a) Schematic and circuit diagram of asymmetric SQUID device and measurement. (b) Measured $I_c(\Phi)$ of an Asymmetric SQUID fabricated on an La$_{1.875}$Ba$_{0.125}$CuO$_4$ crystal for a variety of temperatures.}
    \label{asym}
    \end{figure}
    
    In addition to probing the current-phase relation of junctions on La$_{2-x}$Ba$_x$CuO$_4$ crystals using the flux transformer technique, we used the same crystal fabrication methods to pattern an asymmetric LBCO-Au-Nb SQUID onto the corner of an LBCO crystal at $x=1/8$ doping, allowing us to measure the current-phase relation of the smaller junction (Fig. \ref{asym}a). For a SQUID containing junctions with current-phase relations I$^L(\phi_1)$ and I$^S(\phi_2)$ under an applied field $\Phi$, flux quantization requires $\phi_2-\phi_1=2\pi$$\Phi/\Phi_0$. A current $I$ biased through the loop should then follow $I(\Phi)=I^L(\phi_1)+I^S(\phi_1+2\pi\Phi/\Phi_0)$. For $I_c^L>>I_c^S$, $I_c$ should be reached when $\phi_1\approx\pi/2$, yielding $I_c(\Phi)\approx$$I_c^L+I^S(\pi/2+2\pi\Phi/\Phi_0)$. Hence, the modulation in $I_c$ of the SQUID due to an applied field $\Phi$ near the critical current $I_c^L$ of the large junction represents the current-phase relation $I^S(\phi_2)$ of the smaller junction, or $I^S(\phi_2)\approx$$I_c(\Phi)-I_c^L$ \cite{della_rocca_measurement_2007,nanda_current-phase_2017}.  

\begin{figure}
    \includegraphics[width=\columnwidth]{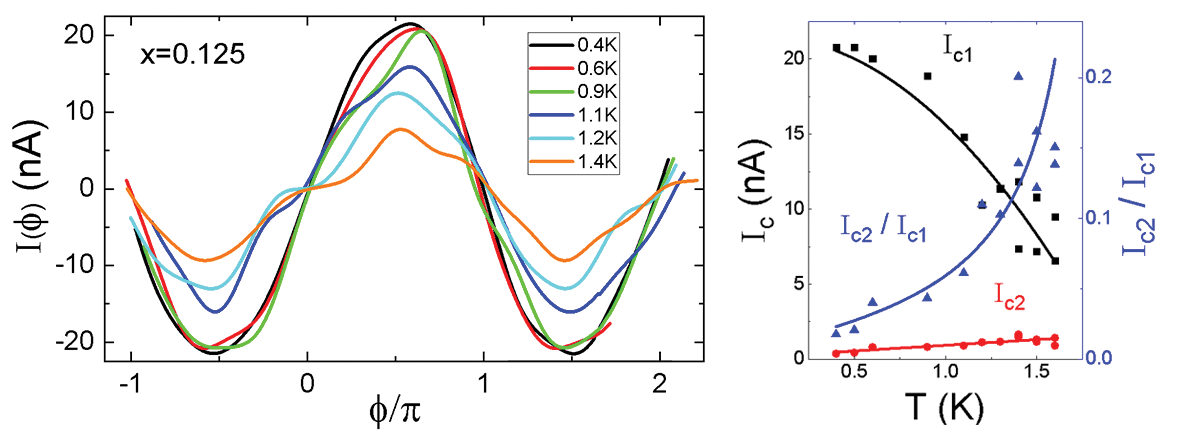}
    \caption{Subtracting $I_c^L$, we obtain the current phase relation $I^S(\phi)$ for a range of temperatures. As T increases, I$_{c1}$ is suppressed and a visible $2^{nd}$ harmonic begins to appear. Fourier analysis of the current-phase relation measurements obtained with the asymmetric SQUID technique shows, as in Fig.\ref{directcpr}e, that the sin(2$\phi$) component of the current-phase relation increases as the typical sin($\phi$) component becomes suppressed.}
    \label{asymcpr}
    \end{figure}

The results of the asymmetric SQUID measurement are shown in Fig. \ref{asym}b for the samples at $x=1/8$ doping. We measured a periodic change in $V(\Phi)$ of the SQUID for a fixed bias current $I>I_c$, and converted this to $I_c(\Phi)$ using the resistively shunted junction model \cite{forder_useful_1977}. We observed a periodic modulation in $I^S$ about an $I_c^L$ which remained roughly constant over the measured field interval for a given $T$. Subtracting $I_c^L$, we obtain the current phase relation $I^S(\phi_2)$ of the smaller junction, plotted in Fig. \ref{asymcpr} for select temperatures. Fourier analysis of this data shows us that $I_{c1}$, the first harmonic of $I^S(\phi_2)$, decreases with temperature, while the second harmonic, $I_{c2}$, increases with temperature. This result is consistent with our measurements of the current-phase relations of $x=1/8$ junctions acquired using the flux transformer method.

By significantly increasing the external field applied to our corner SQUID, we were able to see single junction effects that provide information about the pairing symmetry of the superconductivity in LBCO. In the corner junction geometry, part of the Josephson tunneling in the device occurs through the crystal face aligned to the $a-$direction, and part of it occurs in the $b-$direction. If the LBCO crystal is an $s-$wave superconductor, we expect to see the typical Fraunhofer-like modulation of a uniform junction, $I(\Phi)=I_0|sin(\pi\Phi/\Phi_0)/(\pi\Phi/\Phi_0)|$, with a peak in the supercurrent at zero field.  However, if the pairing symmetry of superconductivity in the crystal is $d_{x^2-y^2}$, a $\pi$ phase shift between the $a-$and $b-$directions will cause destructive interference between the critical currents through the two facets at zero field, as observed in the measurement of $d-$wave pairing symmetry of YBCO \cite{wollman_experimental_1993}.  

\begin{figure}
    \includegraphics[width=\columnwidth]{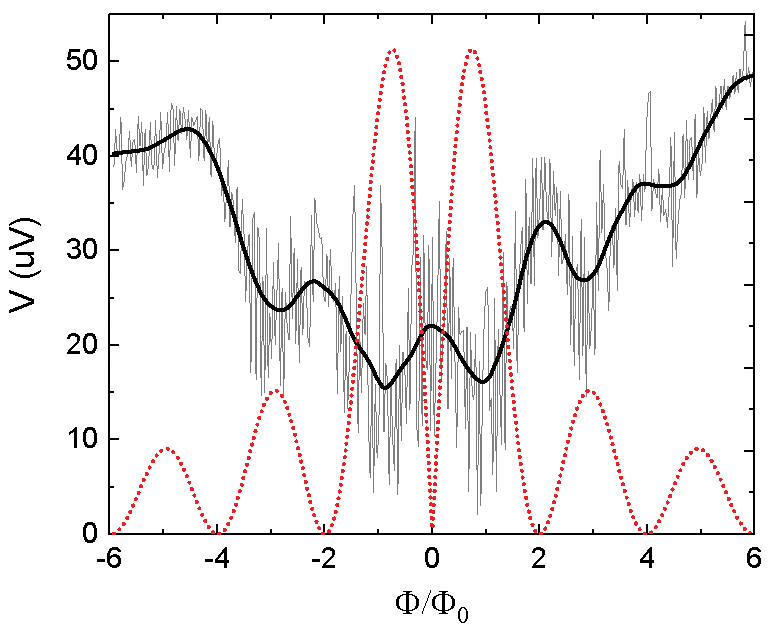}
    \caption{Single junction effects observed in the large field $V(\Phi)$ of a SQUID fabricated on the corner of an La$_{1.875}$Ba$_{0.125}$CuO$_4$ crystal, for fixed bias current $I$. The minima in $V$ (black curve) correspond to the local maxima in the theoretical $I_c(\Phi)$ (red dotted curve) for a junction fabricated on the corner of a crystal with a d-wave pairing symmetry.}
    \label{pairsym}
    \end{figure}

As shown in Fig. \ref{pairsym}, we measured $V(\Phi)$ at a fixed bias current at $T=1K$ and compared it to $I_c(\Phi)$ calculated for an $d_{x^2-y^2}$ corner junction. We would expect the field dependence of a $d-$wave corner junction to follow the functional form $I(\Phi)=I(0)|sin^2(\pi\Phi/2\Phi_0)/(\pi\Phi/\Phi_0)|$, assuming equal lengths of the junction along the $a$ and $b$ directions. Indeed, the measured voltage has a local maximum at zero flux, which corresponds to a local minimum in the critical current at zero flux, as we would expect for a $d_{x^2-y^2}$ superconductor.

In conclusion, we measured the current-phase relation of La$_{1.875}$Ba$_{0.125}$CuO$_4$-Au-Nb crystal SQUIDs using two separate techniques, and we observed the onset of a significant sin($2\phi$)-component in the Josephson CPR which becomes stronger relative to the conventional sin($\phi$)-component as $T$ increases. This phase-sensitive measurement indicates that as $T$ increases, a larger proportion of the superconductivity in the crystal is carried by a state where the superconducting order parameter is spatially modulated, consistent with the PDW state. In the La$_{1.845}$Ba$_{0.155}$CuO$_4$ SQUID we measured, we observed only a relatively small second harmonic of the CPR, which is consistent with the PDW state weakening away from $x=1/8$ doping. Additionally, by increasing the applied field to the corner SQUID device, we observed single junction behavior consistent with a $d_{x^2-y^2}$ pairing symmetry. The observation of the sin($2\phi$) component in the Josephson current and its temperature dependence gives strong support to the proposal of \cite{berg_dynamical_2007} that LBCO at $x=1/8$ harbors a PDW state, in which charge, spin and superconducting order are intertwined.
    
We thank Steve Kivelson for useful discussions, and Juan Atkinson, Erik Huemiller and Adam Weis for technical assistance. This work was funded by the Office of Basic Energy Sciences, Materials Sciences and Engineering Division, U.S. Department of Energy (DOE) under contracts DE$-$SC0012368 at the University of Illinois at Urbana-Champaign and DE$-$SC0012704 at Brookhaven National Laboratory.  The work at the University of Illinois was also supported in part by the NSF grants DMR$-$1710437 (DH) and NSF DMR$-$1725401 (EF).  We also acknowledge the use of the device fabrication and materials characterization facilities of the Materials Research Laboratory at UIUC.

\bibliographystyle{apsrev4-1} 
\bibliography{Zotero.bib} 
\end{document}